\documentclass[useAMS,usenatbib]{mn2e}
\usepackage{graphicx}

\title[Probing the Galaxy's bars via the Hercules stream]{Probing the
  Galaxy's bars via the Hercules stream}

\author[Esko Gardner and Chris Flynn]{Esko
  Gardner$^{1}$\thanks{E-mail:esgard@utu.fi} and Chris Flynn$^{1}$\\
  $^{1}$Tuorla Observatory, Department of Physics and Astronomy,
  University of Turku, V\"ais\"al\"antie 20,FI-21500, Piikki\"o,
  Finland}
  \begin{document}
  \date{April 2010 - corrected version}
  \pagerange{\pageref{firstpage}--\pageref{lastpage}} \pubyear{2010}
  \maketitle
  \label{firstpage}

\begin{abstract}

It has been suggested that a resonance between a rotating bar and
stars in the solar neighbourhood can produce the so called `Hercules
stream'. Recently, a second bar may have been identified in the
Galactic centre, the so called `long bar', which is longer and much
flatter than the traditional Galactic bar, and has a similar mass.  We
looked at the dynamical effects of both bars, separately and together,
on orbits of stars integrated backwards from local position and
velocities, and a model of the Galactic potential which includes the
bars directly. Both bars can produce Hercules like features, and allow
us to measure the rotation rate of the bar(s). We measure a pattern
speed, for both bars, of $1.87 \pm 0.02$ times the local circular
frequency. This is on par with previous measurements for the Galactic
bar, although we do adopt a slightly different Solar motion. Finally,
we identify a new kinematic feature in local velocity space, caused by
the long bar, which is tempting to identify with the high velocity
`Arcturus' stream.

\end{abstract}


\section{Introduction} 

The inner regions of the Milky Way contain a non-axisymmetric
structure sometimes referred to as the triaxial (or boxy) bulge and/or
the Galactic bar. Primary evidence for this structure are the
COBE/Diffyse Infrared Background Experiment (DIRBE) near infrared
(NIR) luminosity maps of the inner Galaxy \citep{Binney97,Bissantz02},
but it has also been mapped via star counts using Two Micron All Sky
Survey (2MASS) \citep{Lopez05}, 2MASS and Optical Gravitational
Lensing Experiment II (OGLE-II) \citep{Vanhollebeke09}. Typical
parameters for the triaxial bulge/bar are axis ratios of 1.0:0.4:0.3,
that it lies at an angle of between 10$^\circ$ and 40$^\circ$ to the
line of sight to the Galactic centre (see e.g. table 1 of
\citealt{Vanhollebeke09}) and that its mass is of order 1 $\times$ 10$^{10}$
M$_\odot$, \citep{Zhao96,Weiner99}.

  It has recently been proposed \citep{Benjamin05} that the inner
Galaxy also contains a `long bar', discovered as an over-density in
star count data at 4.5 $\mu$m taken in the Galactic Legacy Infrared
Mid-Plane Survey Extraordinaire (GLIMPSE) survey of the Galactic
centre made with \textit{Spitzer}. \cite{Benjamin05} measure the
half-length of this bar as $4.4 \pm 0.5$ kpc, and lying at a position
angle of $44 \pm 10^\circ$ to the line of sight to the Galactic
centre. Only the closer end of the bar was detected, as the far end
would lie at apparent magnitudes fainter than the completeness limit
of the survey. The long bar has been confirmed by \cite{Lopez07}, who
detect it via red clump stars identified from the 2MASS survey. They
find a half-length of 3.9 kpc and a position angle of $43 \pm
7^\circ$, very similar to \cite{Benjamin05}.  The long bar is much
longer than any previously proposed bar/bulge in the Galactic centre,
for which the long axis is thought to lie in the range 2.4 (e.g.
\citealt{Dwek95}) to 3.5 kpc (e.g. \citealt{Bissantz02}). Most
interestingly, the long bar seems to lie at a different angle to the
traditional bar/bulge (hereafter `Galactic bar'), for which estimates
typically lie in the range $25 \pm 10$ degrees \citep{Vanhollebeke09}.

  Most recently, \cite{Cabrera07} have used wider field 2MASS data to
isolate red clump stars over a wide range of Galactic longitude,
showing that the long bar, the triaxial bulge and the disc can all be
seen as separate structures (at least on our side of the Galaxy) with
the important result that the long bar is much flatter than the bulge.
The long bar is found to have a vertical scale-height of around 100 pc
whereas the bulge has a vertical scale-height of 500 pc. Furthermore,
the red clump stars used to trace the long bar can be used to estimate
its total mass (assuming that the mass to light ratio of red clump
stars is the same in the long bar as it is in the bulge) as 6 $\times$ 10$^9$
M$_\odot$, of the same order of the mass of the Galactic bar 1 $\times$
10$^{10}$ M$_\odot$, \citep{Zhao96,Weiner99}.

  \cite{Dehnen2000} has shown that a bar in the inner galaxy has
interesting dynamical consequences for the orbits of stars even in the
solar neighbourhood, 8 kpc from the centre, due to a resonance of
stellar orbits with the Outer Lindblad Resonance (OLR) of the bar. The
manifestation of this resonance in our local space is the $\zeta$
Herculis stream \citep{Eggen71}, or the `Hercules stream' as it is
usually called. 

   In this paper, we use a similar method to \cite{Dehnen2000}, to reexamine
the dynamical effects on local disc stars of these non-axisymmetric features in
the Galactic centre. We have examined both the long bar (see \citealt{Lopez07})
and the Galactic bar (see \citealt{Bissantz02}), separately, and as a dual-bar
system.  We adopt R$_\odot = 8$ kpc throughout the paper. This facilitates
comparison with \cite{Dehnen2000}, who adopted the same value.

  We have made a probability-statistical study of the effects of the
bar(s) on local velocity space, by the means of potential theory. We will
present the statistical evidence produced by various variations of the long
bar-system, and a the Galactic bar-system, as well as, some interesting
combined models.

\section{The Hercules stream}  

\begin{figure}
      \begin{center}
	  \includegraphics[width=3in]{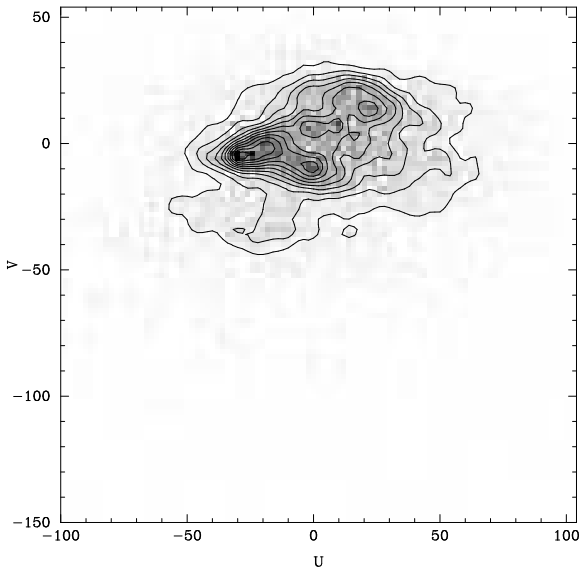}
      \end{center}
      \caption{Local velocity space, data includes all the stars with
	$U$ and $V$ velocities from the Geneva-Copenhagen survey
	\protect\citep{Holmberg07} and \protect\cite{Schuster06}. 
        The Hercules stream is the prominent
	feature area around ($-25$,$-30$) km s$^{-1}$. The velocities 
	have been corrected to the solar velocity from 
	\protect\cite{Schonrich10}}\label{gcs}
     \end{figure}

  In Fig. \ref{gcs} we show local velocity space in $(U,V)$, where $U$
is the velocity towards the Galactic centre and $V$ is the velocity in
the direction of Galactic rotation. The figure has been prepared using
12939 stars from the Geneva Copenhagen Survey \citep{Holmberg07} of
abundances, ages and kinematics of nearby stars, and has been
supplemented by 1535 stars from the the high velocity catalogue of
\cite{Schuster06}. The bin size is 2 km s$^{-1}$, and the contours are
in 5 percent steps relative to the maximum density. The velocities
have been corrected for the solar motion taken from \cite{Schonrich10}
of $(U_\odot,V_\odot) = (11.1,12.2)$ km s$^{-1}$. The Hercules stream
is the feature at approximately $(U,V) = (-25,-30)$ km s$^{-1}$. This
is the feature which \cite{Dehnen2000} identifies as having a
dynamical origin as a resonance with the bar.  Our simulations in this
paper of a Galactic bar, long bar (and both) are primarily constrained
by this feature.

The bulk of the stars in Fig. \ref{gcs} belong to the complicated
feature above the Hercules stream.  The classically identified
features in this region are the Pleiades, Sirius and Hyades streams,
which occur at around $(0,-10)$, $(20,20)$, and $(-25,-5)$ km s$^{-1}$
respectively. Further streams than those mentioned have been
identified by a number of groups, but these mainly concern higher
velocity stars and cannot be pointed to easily in Fig \ref{gcs}. One
of the very interesting features at higher velocities is the Arcturus
stream at $V \approx 100$ km s$^{-1}$ (for a wide range of $U$
velocities) identified by \cite{Eggen71-2} and discussed recently by
\cite{Williams09}, who have identified it in the RAdial Velocity
Experiment (RAVE) survey. The Arcturus stream may have a dynamical
origin, as we show using the simulations in this paper.

\section{Modelling the disc and bar}

  We model the effects of a bar on local disc stars by setting up a
potential describing the disc, bulge, dark halo and bar, and
integrating a library of orbits passing through the Solar neighbourhood,
similarly to \cite{Dehnen2000}.

  We begin with the axisymmetric Galactic potential of \cite{Flynn96}.
This model contains a disc, bulge and dark halo. The scale-length of
disc matter ($h_R$) in the model is 4.4 kpc, which these days appears
rather long, as recent studies of the disc scale-length give results
closer to 3 kpc. Analysis of NIR data in J and K with DEep Near
Infrared Survey of the Southern Sky (DENIS) give $h_R = 2.3 \pm 0.1$
kpc \citep{Ruphy96}, while K giants seen in 2MASS give $h_R = 3.3 \pm
0.1$ kpc \citep{Lopez02}. These surveys tend to probe the scale-length
of luminosity in the disc, rather than the mass, which is traced
rather better via main sequence dwarfs. \textit{Hubble Space Telescope
  (HST)} star counts of M dwarfs
yield a disc scale-length of $h_R = 3.0 \pm 0.4$ \citep{Gould97}. F
and K dwarfs, traced in huge numbers in the 6500 deg$^2$ SDSS
survey, yield a disc scale-length of $h_R = 2.6 \pm 0.5$
\citep{Juric08}. For a more complete review of the disc scale-length,
see \cite{Gardner08}. We chose a disc scale-length of 3 kpc to reflect
the combined observational constraints. This was straightforward and
only involved modifying the Miyamoto discs from which the
exponential-like disc is built up.

\begin{figure}
 \includegraphics[width=\columnwidth]{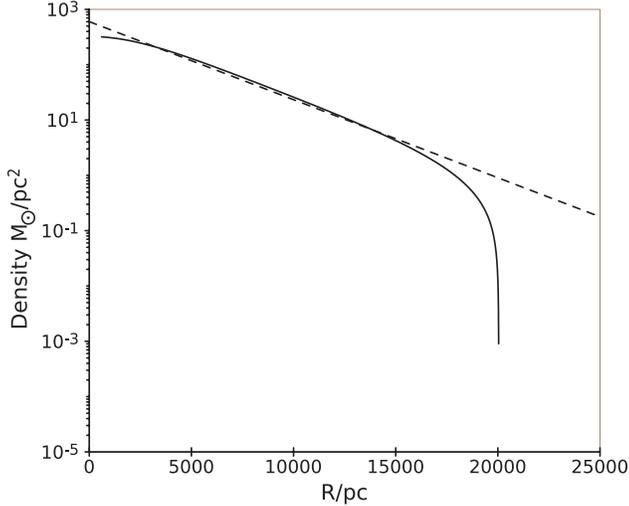} 
  \caption{The surface density of the disc component (solid line) as a function of
    Galactocentric radius. The dashed line corresponds to an
    exponential density falloff, 3 kpc, which is a good fit to the
    model over a wide range of radii. Note that the density truncates
    strongly at 18 kpc.}\label{discdensity}
\end{figure}

  The surface density ($\Sigma_\odot$) of the new disc at the Sun's position is
50 M$_\odot$ pc$^{-2}$ and the local density $\rho_\odot$ is 0.10 M$_\odot$
pc$^{-3}$, consistent with observational constraints \citep{Holmberg00}. The
surface density of the disc as a function of Galactocentric radius is shown in
Fig. \ref{discdensity}. The new disc model has a 3 kpc scale-length, with a
truncation at about 18 kpc.

The full parameters of the new disc model are shown in Table \ref{modelparams},
along with the other parameters of the model (which are unchanged but are
reproduced for clarity).

The equations for the axisymmetric potential are: \\

  \begin{table}
   \begin{center}
      \caption{Parameters of the model.}\label{modelparams}
      \begin{tabular}{lcr}
	  \hline
	  Parameter & Value & Unit\\
	  \hline
	  V$_h$ & 220 & km s$^{-1}$\\
	  r$_0$ & 8.5 & kpc\\
	  \hline
	  $M_{C_1}$ & 3 & 10$^9$ M$_\odot$\\
	  $r_{C_1}$ & 2.7 & kpc\\
	  $M_{C_2}$ & 16 & 10$^{9}$ M$_\odot$\\
	  $r_{C_2}$ & 0.42 & kpc\\
	  \hline
	  M$_{d_1}$ & 77.04 & 10$^9$ M$_\odot$\\
	  a$_{d_1}$ & 5.81 & kpc\\
	  M$_{d_2}$ & $-$68.48 & 10$^9$ M$_\odot$\\
	  a$_{d_2}$ & 17.43 & kpc\\
	  M$_{d_3}$ & 26.75 & 10$^9$ M$_\odot$\\
	  a$_{d_3}$ & 34.84 & kpc\\
	  b & 0.3 & kpc \\
	  \hline
      \end{tabular}
      \end{center}
  \end{table}

  \begin{equation}
      \Phi = \Phi_H + \Phi_C + \Phi_D + \Phi_{bar}
  \end{equation}  
  \begin{equation}
      \Phi_H = {1\over{2}}V_h^2 {\rm ln}(r^2 + r_0^2)
  \end{equation}  
  \begin{equation}
      \Phi_C = - {{GM_{C_1}}\over{\sqrt{r^2+r^2_{C_1}}}} - {{GM_{C_2}}\over{\sqrt{r^2+r^2_{C_2}}}}
  \end{equation}  
  \begin{equation}
      \Phi_{D_n} =
      {-GM_{D_n}\over{\sqrt{(R^2+(a_{d_n}+\sqrt{(z^2+b^2)})^2)}}} ,
  \end{equation}
where $\Phi_D=\Sigma_{n=1}^3 \Phi_{D_n}$, $r=x^2+y^2+z^2$, and
$R=x^2+y^2$.

We have used a different method to modelling the bar as was adopted in
\cite{Dehnen2000} and \cite{Minchev07}. Instead of modelling the bar
as a local quadrupole perturbation to the potential, we model the bar
directly in the inner galaxy. This means
that orbits of stars entering into the inner disc regions and even
passing through the bar are much more accurately modelled, even if it
is computationally far more expensive.

The bar is modelled as a Ferrers $n=2$ potential, as laid out by
\cite{Pfenniger84}, and as such, is a triaxial ellipsoid. The
equations for the bar are too complex to present here (for full
details see \citealt{Pfenniger84}). The bar rotates as a rigid body.

A bar is specified by its three axis lengths, its mass, its angular
(or pattern) speed and its angle to the line connecting the Galactic
centre and the Sun.

We have two standard bars, with parameters taken from \cite{Lopez07} and from
\cite{Bissantz02}. The dimensions of the `long bar' are (3.9:0.6:0.1) kpc, with
a mass of 6 10$^9$ M$_\odot$ and an angle of 43$^\circ$. The `Galactic bar' has
dimensions of (3.5:1.4:1.0) kpc, a mass of 10$^{10}$ M$_\odot$ and an angle of
25$^\circ$. These masses correspond to a \cite{Dehnen2000} $\alpha$ value of
0.0037, for the long bar case, and 0.0040, for the Galactic bar case. Note that
in the simulations we do not vary the dimensions of the respective bars, only
the mass and position angle.
\begin{table}
    \begin{center}
	\caption{Default simulation parameters.}\label{simparams}
	\begin{tabular}{llcr}
	    \hline
 	    Parameter & Value & Unit\\
	    \hline
	    \textbf{Galactic Bar}& &\\
	    Angle & 25 & $^\circ$\\
	    Dimensions & 3.5:1.4:1.0 & kpc\\
	    Mass & 10 & 10$^9$ M$_\odot$\\
	    Pattern Speed & 55.9 &km s$^{-1}$ kpc$^{-1}$\\
	    Local standard of rest & 239 & km s$^{-1}$\\
	    OLR & 1.87 &\\	    
	    \hline
	    \textbf{Long Bar} & &\\
	    Angle & 43 & $^\circ$\\
	    Dimensions & 3.9:0.6:0.1 & kpc\\
	    Mass & 6 & 10$^9$ M$_\odot$\\
	    Pattern Speed & 54.9 &km s$^{-1}$ kpc$^{-1}$\\
	    Local standard of rest & 235 & km s$^{-1}$\\
	    OLR & 1.87 &\\
	    \hline
	\end{tabular}
    \end{center}
\end{table}
  
An important parameter is the pattern speed of the bar. Following
\cite{Dehnen2000} and \cite{Minchev07}, we parametrise the bar's pattern speed
by the OLR, i.e. by the ratio of the period of a
star with velocity of the local standard of rest (LSR) in a given Galactic
model to the period of rotation of the bar.

The definition of the LSR is straightforward in axisymmetric models of
the Galaxy, but needs to be redefined slightly once we have added a
bar to the model. In the presence of a bar, a star can no longer
follow a circular orbit at the solar radius, but rather oscillates
around a mean radius. For typical bar parameters, the oscillation is
up to 10 km s$^{-1}$, compared to a mean velocity of 220 km s$^{-1}$
in typical models. For each model run in our simulations, we
determine, experimentally, at what velocity a star, which starts at
the Solar position, would have a mean orbital radius of 8 kpc.
Integration is carried out for 10 Gyr, so that the effect of the
effect of the angle of the bar is averaged out. This is defined as the
LSR for that model.

\section{Simulations and methods}

The basic simulation approach is to model the appearance of local velocity
space using the method described by \cite{Dehnen2000}, a `backward restricted
$N$-body method'. The idea is to trace a library of orbits starting from the
solar neighbourhood backwards over some fixed time (typically 1 Gyr), and use
the known density distribution and velocity distribution of the disc to 
reconstruct the velocity distribution of stars in the solar neighbourhood.

We introduce two refinements, firstly, a more realistic representation
of the bar potential and secondly, we use observational constraints on
the velocity dispersion and asymmetric drift of stars in the Galactic
disc over a wide range of Galactocentric radius obtained by
\cite{Lewis89}. In practise, this latter refinement is not very
different to the approach adopted by \cite{Dehnen2000}, (simple
exponential functions of stellar density and radial velocity
dispersion for the old stellar disc) and indeed we find that we
reproduce his results quite well when we adopt a bar similar to his.
Our simulations are in two dimensions (2D), as in \cite{Dehnen2000}.

The initial conditions are set by a velocity-grid, of 100 by 100 elements in
the $UV$-velocity plane, covering $U = -100$ to $100$ km s$^{-1}$ , and V$=-$150
km s$^{-1}$) to $V= 50$ km s$^{-1}$), in steps of 2 km s$^{-1}$. A tracer with the
$UV$-velocity selected from the grid is then set into motion, following the
tracer backwards in time over a period of 1 Gyr, tracking its position every 1
Myr. The initial position is selected as the solar position (8 kpc), and both
the $z$-coordinate and the $W$-velocity are set to zero. The initial velocity in
the rotational direction $V$ is, of course, relative to the LSR for particular model
being tested. 

Statistically, we use a density and velocity-based system, where a weight value
corresponds to a 8 kpc-centric exponential distribution, and a Gaussian
velocity-distribution, with standard deviations as described in
\cite{Lewis89}. The $V$-velocity is corrected back to a `true' $V$-velocity using
the LSR of the position and the asymmetric drift (\citealt{Lewis89}).\\
  
\section{Results}

We examine in this section the effects of both the `Galactic bar' and the `long
bar' on local velocity space. We vary the bar mass, value of the OLR (i.e. the
pattern speed of the bar), and the angle of the bar. Finally, we look at the
effects of including both bars at the same time.

\subsection{Effect of the mass of the Galactic bar}

\begin{figure}
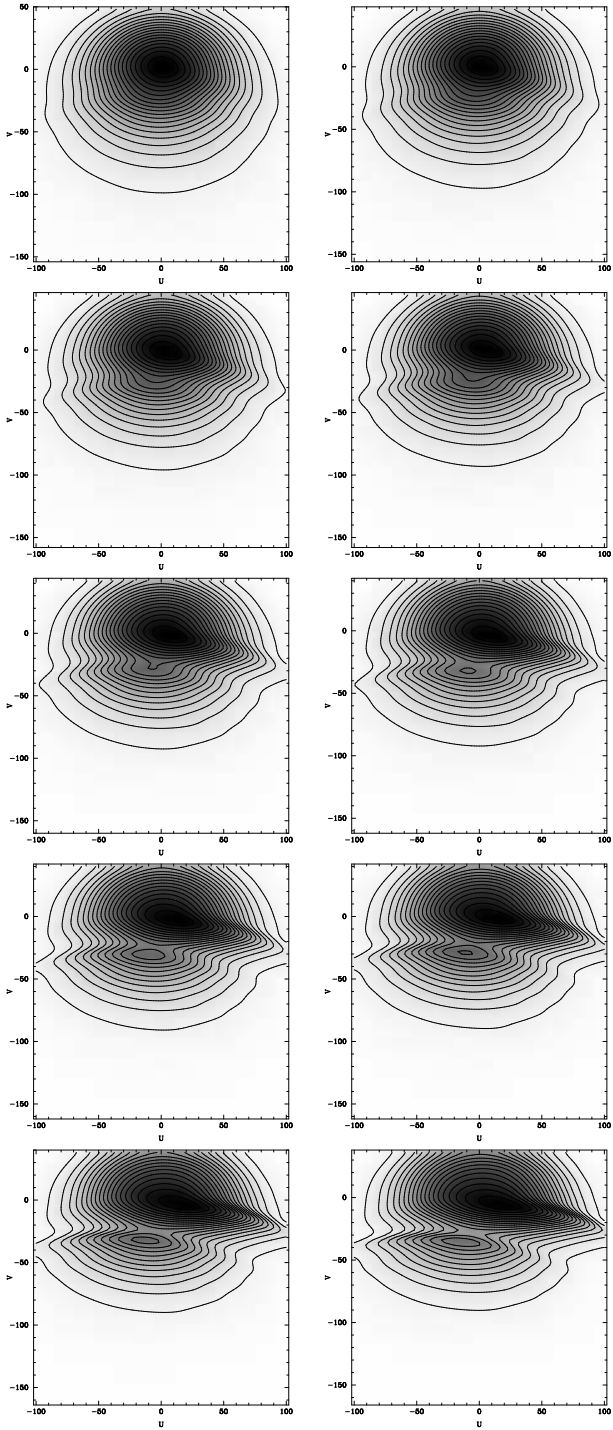

  \begin{tabular}{ll}
      \includegraphics[width=1.5in]{cobe-150-1e9.eps} &
      \includegraphics[width=1.5in]{cobe-150-2e9.eps} \\
      \includegraphics[width=1.5in]{cobe-150-3e9.eps} &
      \includegraphics[width=1.5in]{cobe-150-4e9.eps} \\
      \includegraphics[width=1.5in]{cobe-150-5e9.eps} &
      \includegraphics[width=1.5in]{cobe-150-6e9.eps} \\
      \includegraphics[width=1.5in]{cobe-150-7e9.eps} &
      \includegraphics[width=1.5in]{cobe-150-8e9.eps} \\
      \includegraphics[width=1.5in]{cobe-150-9e9.eps} &
      \includegraphics[width=1.5in]{cobe-150-10e9.eps} \\
  \end{tabular}
  \caption{Effect on local velocity space 
    of varying the mass of the Galactic bar, from 10$^9$ to 10$̂^{10}$
    M$_\odot$ in increments of 10$^9$ M$_\odot$. For other simulation
    parameters see Table \ref{simparams}.}\label{cobemass}
  \end{figure}

We begin by running the standard model of the `Galactic bar' but
varying its mass over a range of 1 to 10 $\times$ 10$^9$ M$_\odot$,
showing the results in Fig. \ref{cobemass}. The OLR for this case is
1.87. The variations of mass within one order of magnitude have
prominent effects. As found by \cite{Dehnen2000} we find that a
Hercules-like feature arises in local velocity space in the presence
of a bar. We will henceforth refer to this as a secondary feature in
velocity space, with the primary feature (above it) containing most of
the stars. The effect of increasing the mass of the bar is similar to
that found by \cite{Dehnen2000}, with the strength of the secondary
feature increasing (and the depth of the resonance gap between the
secondary feature and primary feature also growing). The secondary
component becomes more ellipsoidal, as mass is increased, and, as
expected, becomes more prominent. Also, the position of the secondary
feature moves to more negative $V$ velocities as the mass is
increased. While it is clear that something like the Hercules feature
can be generated by this bar, the detailed match is not yet perfect.
In Fig. \ref{gcs}, the Hercules feature is found at $(U,V) =
(-25,-30)$ km s$^{-1}$. In the simulations, the generated feature
approaches this location for the higher mass models, but at the
expense of generating a sharp `tail' of stars at positive $U$ values
spreading outward from the primary feature. This tail of stars is also
seen in the \cite{Dehnen2000} and \cite{Minchev07} simulations, and
increases with the mass of the adopted bar. Such a tail, if present at
all in the observational data (Fig. \ref{gcs}) is quite weak. Apart
from this tail, the generated feature is fairly consistent with
various estimates of bar mass where they are consistently around and
over 10$^{10}$ M$_\odot$ (see e.g. \citealt{Zhao96,Weiner99}).

\subsection{Long bar}

\begin{figure}
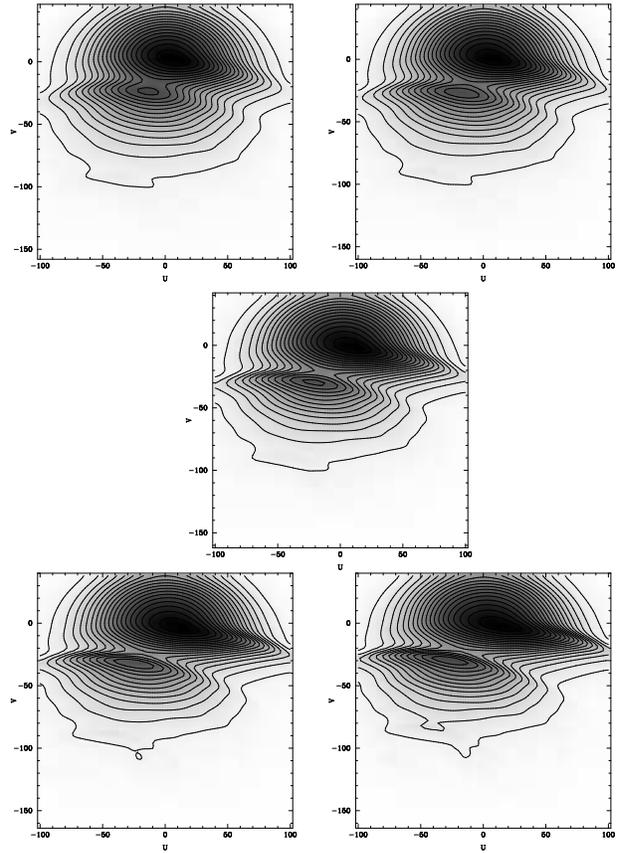

\begin{tabular}{ll}
      \includegraphics[width=1.5in]{long-n2-4over6times-187olr-backwards.eps} &
      \includegraphics[width=1.5in]{long-n2-5over6times-187olr-backwards.eps}\\      
\end{tabular}
  
\hspace{1in}\includegraphics[width=1.5in]{long-n2-187olr-backwards.eps} 

\begin{tabular}{ll}
      \includegraphics[width=1.5in]{long-n2-7over6times-187olr-backwards.eps} &
      \includegraphics[width=1.5in]{long-n2-8over6times-187olr-backwards.eps} \\
\end{tabular}
\caption{Variations of velocity space with different values of
    for long bar mass. Masses are $4$, $5$, $6$ (middle), $7$, and $8$ $10^9 
    M_\odot$.  For other simulation
    parameters see Table \ref{simparams}.}\label{barmass}
\end{figure}

We next look at the `long bar' case. We adopt the standard model for
its dimensions and initial position angle and vary only the mass.
These models all give a Hercules-like secondary feature, similar to
the Galactic bar case. The effect of varying its mass on local
velocity space is shown in Fig. \ref{barmass}. One can see that mass
takes a significant role in the division of the primary and secondary
feature, where the secondary is stronger with increasing mass. Again,
these plots superficially resemble the real local velocity space (Fig.
\ref{gcs}) but not in detail. The position of the secondary feature
tends to be at too high $V$ velocities and the tail of stars at high
$U$ values also appears for the higher mass models. 

One very interesting `tertiary' feature appears with the long bar, located at
$V = -80$ to $-100$ km s$^{-1}$. The tertiary feature becomes more prominent as
mass increases. This feature is not seen in the Galactic bar simulations
(Fig. \ref{cobemass}). We will return to this feature in section \ref{anglesection}.

\subsection{Pattern speed of the bar}\label{olrsection}

As shown by \cite{Dehnen2000} and \cite{Minchev07}, the position of the
Hercules feature is quite dependent on the adopted pattern speed of the bar,
parametrised by the OLR. The feature can therefore be used to measure the
pattern speed of the bar. \cite{Dehnen2000} finds that the feature can be fit
by an OLR of 1.85 $\pm$ 0.15, using similar simulations to
ours. \cite{Minchev07} find a similar and more tightly constrained value of
the OLR of 1.87 $\pm$ 0.04, using similar simulations as here but also from
constraints implied by the locally measured Oort constant $C$.

\begin{figure}
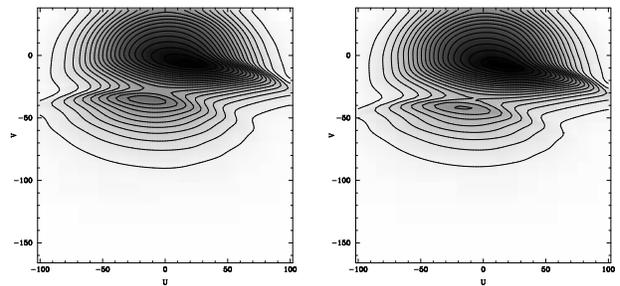

    \begin{tabular}{ll}
  	\includegraphics[width=1.5in]{cobe-150-10e9.eps} &
	\includegraphics[width=1.5in]{cobe-10e9-190olr.eps} \\
    \end{tabular}
    \caption{Variation of velocity space with a Galactic bar model,
      OLR values of 1.87 and 1.90.}\label{olrcobe}
\end{figure}

\begin{figure}
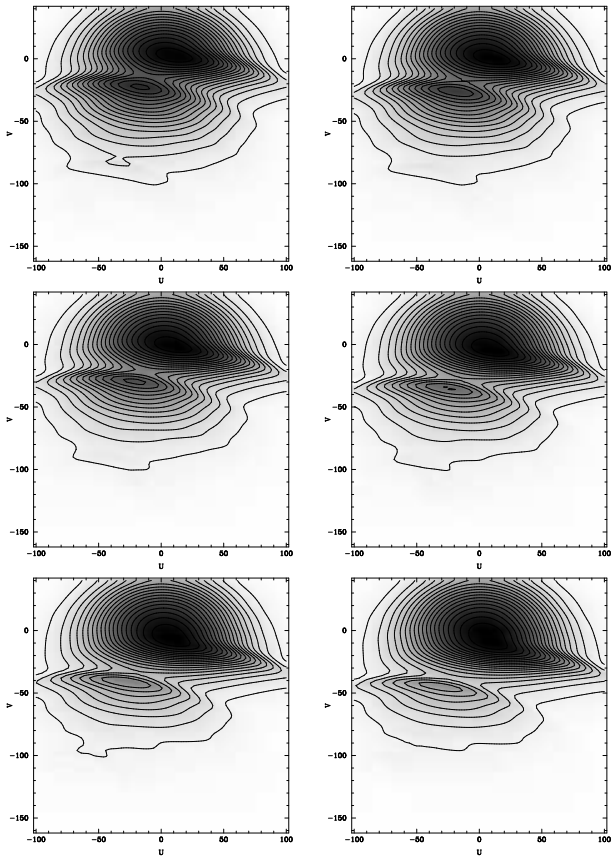

  \begin{tabular}{ll}
      \includegraphics[width=1.5in]{long-n2-183olr-backwards.eps} &
      \includegraphics[width=1.5in]{long-n2-backwards.eps}\\
      \includegraphics[width=1.5in]{long-n2-187olr-backwards.eps} &
      \includegraphics[width=1.5in]{long-n2-190olr-backwards.eps} \\
      \includegraphics[width=1.5in]{long-n2-193olr-backwards.eps} &
      \includegraphics[width=1.5in]{long-n2-195olr-backwards.eps}
  \end{tabular}
  \caption{Effect on local velocity space with different values of
    for the OLR for the long bar case. Values are 1.83, 1.85, 1.87, 1.90,
    1.93, and 1.95, from left to right, top to bottom. There is a clear shift
    in the position of the secondary feature to lower $V$ velocity as the 
    OLR increases, allowing a determination to be made of the pattern 
    speed of the bar. }\label{olrlong}
  \end{figure}

Our standard model for both the Galactic and long bars assumes an OLR of 1.87.
For the Galactic bar case, the effect of shifting from an OLR value of 1.87 to
1.90 is shown in Fig \ref{olrcobe}. The secondary feature shifts by about 5 km
s$^{-1}$ to more negative $V$ velocities. A more detailed view of this is shown
for the long bar case in Fig. \ref{olrlong}, where we have varied the value of
the OLR ranging from 1.83 to 1.95. The secondary feature shifts quite markedly
away from the primary feature over this range of assumed OLR values. We derive
a best fitting OLR of $1.87 \pm 0.02$, assuming that the position of the
secondary feature is at $V = -30 \pm 5$ km s$^{-1}$. 

Our determination of the OLR value for the bar is in agreement with
\cite{Dehnen2000} and \cite{Minchev07} within the error bars. We note that our
value about the same as theirs, even though we adopt the solar
motion of \cite{Schonrich10}, which shifts this feature by 7 km s$^{-1}$ to
more positive $V$ velocities in the observational plane (the Hercules feature
is centred on about $V = -35$ km s $^{-1}$ with the solar motion adopted by
\cite{Dehnen2000}, rather than at $V = -30$ km s $^{-1}$ as in our
Fig. \ref{gcs}. Had we adopted the older solar motion, we would obtain an OLR
of $1.90 \pm 0.03$.

\subsection{Bar angle}\label{anglesection}

\begin{figure}
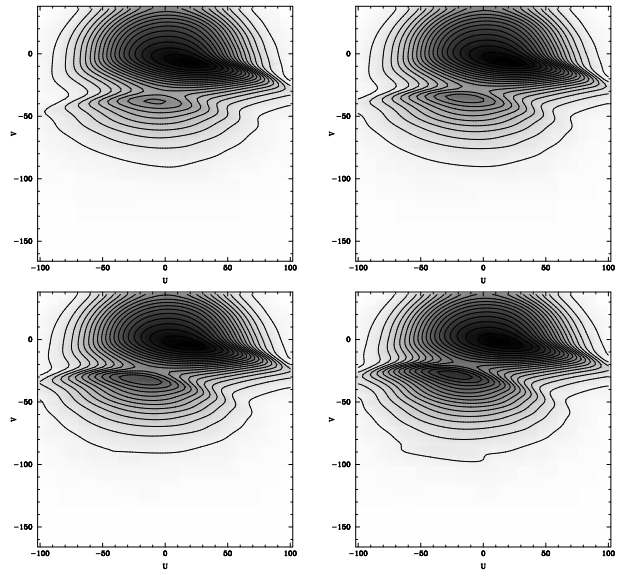

    \begin{tabular}{ll}
	\includegraphics[width=1.5in]{cobe-10e9-15deg.eps} &
	\includegraphics[width=1.5in]{cobe-150-10e9.eps} \\
	\includegraphics[width=1.5in]{cobe-10e9-35deg.eps} &
	\includegraphics[width=1.5in]{cobe-10e9-45deg.eps}
    \end{tabular}
    \caption{Variation of Galactic bar angle, 15$^\circ$, 25$^\circ$, 35$^\circ$, and
      45$^\circ$}\label{cobeangle}
\end{figure}

\begin{figure}
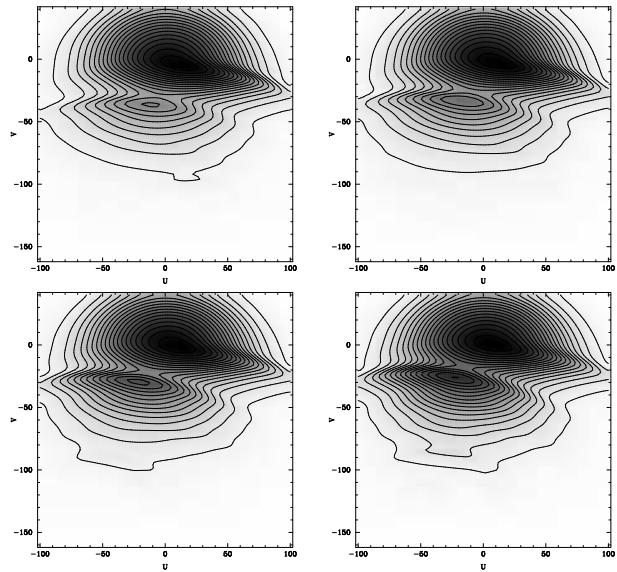

    \begin{tabular}{ll}
	\includegraphics[width=1.5in]{long-n2-23deg-187olr-backwards.eps} & 
	\includegraphics[width=1.5in]{long-n2-33deg-187olr-backwards.eps} \\
	\includegraphics[width=1.5in]{long-n2-187olr-backwards.eps} &
	\includegraphics[width=1.5in]{long-n2-53deg-187olr-backwards.eps}
    \end{tabular}
    \caption{Variation of long bar angle, 23$^\circ$, 33$^\circ$, 43$^\circ$, and
      53$^\circ$.}\label{longangle}
\end{figure}

\cite{Dehnen2000} used the Hercules stream to probe the angle of the
bar, deriving an angle of 25$^\circ$. This is to be compared to the
observational evidence for the (Galactic bar) angle, which lies in the
range 20$^\circ$ to 40$^\circ$. We probe the effect of bar angle as
well, for both the Galactic bar and long bar cases. As found by
\cite{Dehnen2000}, the effect is to change of the position of the
secondary feature to towards more negative values of $U$ and more
positive values of $V$ with increasing bar angle, and also to change
the strength of the feature, as seen in Figs \ref{cobeangle} and Fig.
\ref{longangle}. However, these changes are rather subtle and we
consider them too weak to constrain the bar angle.

The tertiary feature is clearly dependant on bar angle, both for the long bar
case and the Galactic bar case. The feature appears in both cases when the
initial bar angle is greater than 40$^\circ$. The orbits of stars in this
feature are highly eccentric and take them to within 2 kpc of the Galactic
centre, ideal for direct interaction with a bar. This accounts for why bar
angle is important to the generation of the tertiary feature.

There is a stream in the Solar neighbourhood at about this velocity,
$V \sim -100$ km s$^{-1}$, the Arcturus stream \citep{Eggen71-2}. The
stream has recently been recovered in the RAVE survey by
\cite{Williams09}. \cite{Williams09} argue against an accretion origin
for this stream, because its stars have a wide range of metallicity,
so that the accreted system would have to have an unreasonably high
mass, perhaps as large as the Large Magellanic Cloud (LMC). It is
interesting that our simulations show that a dynamical origin for the
stream is plausible. Streams with a dynamical origin are expected to
have a wide range of metallicity, as wide as the stars of its parent
population. If so, the more plausible bar for originating this stream
is the long bar, because its measured bar angle is 43$^\circ$. The
Galactic bar could also produce this stream, but it would have to also
be at an angle of about 45$^\circ$, whereas most studies find that it
lies much closer to the line of sight to the Galactic centre, at about
20$^\circ$ \citep{Vanhollebeke09}.

\subsection{Integration time}

\begin{figure}
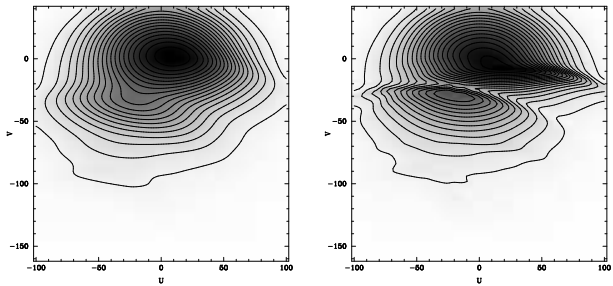

    \begin{tabular}{ll}
	\includegraphics[width=1.5in]{long-150-halftime-187.eps} &
	\includegraphics[width=1.5in]{long-150-doubletime-187.eps}
    \end{tabular}
    \caption{Variation in integration time, 500 Myr, and 2 Gyr, using
      the long bar.}\label{timedomain}
\end{figure}

We have chosen and integration time of 1 Gyr, but this choice is somewhat
arbitrary.  Fig. \ref{timedomain} shows the effects of adopting shorter and
longer integration times of 500 Myr and 2 Gyr for the long bar.

After 500 Myr, the secondary feature has started forming but is not yet
prominent. This represents about 4 bar rotations, and it is not surprising that
it takes some time for the resonance with the bar to develop the trough between
the primary and secondary features. After 2 Gyr we can see the stronger
splitting of the primary and secondary feature, with the resonance-area
clearing, as well as the primary feature tailing out to more positive $U$.
Simulations as long as this assume of course that the bar has existed for 2 Gyr
with the same parameters, in particular that the rotation rate and mass have
not changed during this time. Our choice of 1 Gyr is meant to allow structure
enough time to develop, while not taxing the credibility of bar
properties being very
long lived in galaxies like the Milky Way. \cite{Fux01} has pointed
out a potential difficulty with the backwards integration technique
used here. He finds that as one integrates further backwards in time,
increasingly spurious structures develop in the $UV$-plane. We find no
such behaviour over approximately 16 periods of the bar (Fig.
\ref{timedomain}). We are confident in our 1 Gyr timescale, for
interpreting the simulations.

\subsection{Dual bar models}

  Our simulations have shown that both the Galactic bar, and the newly
discovered `long bar' can generate features in velocity space which are
superficially similar to those seen in the Solar neighbourhood, in particular
the Hercules stream. We now investigate the effect of including both bars.

\begin{figure}
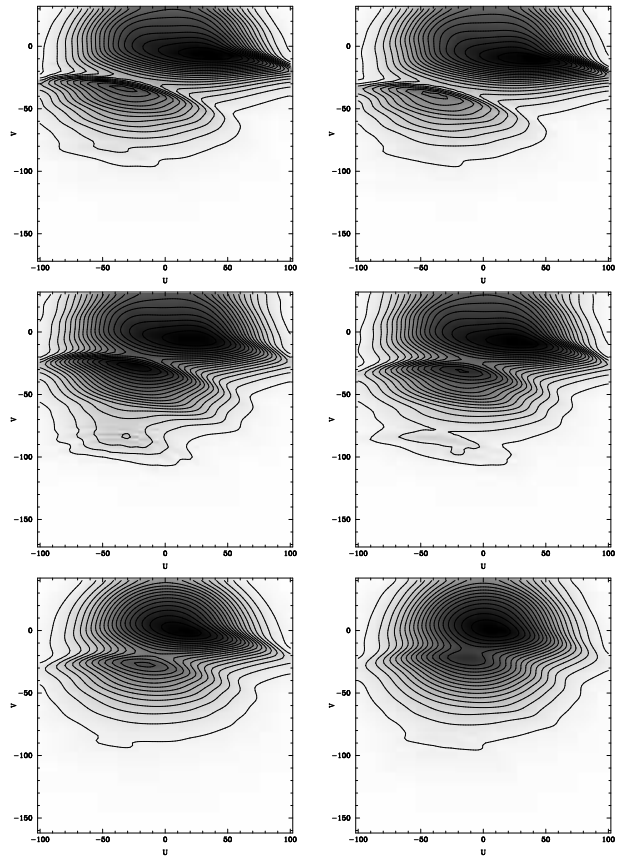
 \begin{tabular}{ll}

\includegraphics[width=1.5in]{dual-150-187l.eps} &
\includegraphics[width=1.5in]{dual-150-190l.eps}\\
\includegraphics[width=1.5in]{dual-150-187-c150.eps} &
\includegraphics[width=1.5in]{dual-150-c187-150.eps} \\
\includegraphics[width=1.5in]{dual-150-half-187l.eps} &
\includegraphics[width=1.5in]{dual-150-half-187l-150c.eps}\\
	      
  \end{tabular}
  \caption{Dual bar models, phase-locked 1.87 and 1.90 (top),
    unlocked, long at 1.87, galactic at 1.5 (middle left), long at 1.5
    galactic at 1.87 (middle right), half-mass locked at 1.87 (bottom
    left), half-mass unlocked long at 1.87, galactic at 1.5 (bottom
    right).}\label{dual-bars}
  \end{figure}

  We show a selection of dual bar models in Fig. \ref{dual-bars}. We begin with
both bars having the same OLR - i.e. the two bars have the same pattern speed
and rotate together. The upper panels of Fig. \ref{dual-bars} show the local
effect on phase space for OLR values of 1.87 and 1.90. The overwhelming
impression here is that the main feature in phase space is driven to large
values of $U$, inconsistent with observations. The main cause for this is
simply the large amount of mass which now resides in the Galactic centre, in
these cases 1.6 $10^{10}$ M$_\odot$. Similar behaviour is seen even for cases
with a single bar, when the mass of the bar exceeds about $10^{10}$ M$_\odot$.
More than this much mass and the bar(s) starts to generate an unacceptable
amount of activity on the Solar neighbourhood $(U,V)$ distribution, typically
pushing stars to too large values of $U$. 

  Two decoupled scenarios were tested, with one of the bars at an OLR
of 1.50 and the other at 1.87. These are shown in the middle panels of
Fig. \ref{dual-bars}.  On the right, the Galactic bar has an OLR of
1.50, while the long bar has an OLR of 1.50 on the left. Again, the
total mass in the two bars is generating significant
structure in local velocity space, with the same problem of too many
stars being shifted into a tail of high $U$ velocities. In addition,
the tertiary feature at $V \sim -80$ becomes very prominent, as the
chances of an interaction with either bar is now higher. We tested
this by running single-bar models with an OLR of 1.5, which results in no
secondary or tertiary feature for the Galactic bar case, and in the long
bar case it causes no secondary and a very weak tertiary, such as seen
in other single-bar simulations of the long bar. Both the upper and
middle panels show more structure than really exists in the Solar
neighbourhood, primarily because too much mass is assigned to the bar
structures in the inner Galaxy. To alleviate this problem, we tested a
scenario where the masses of both bars are simply halved. This is
shown in the lower panels of Fig. \ref{dual-bars}. On the left, the
bars both have an OLR of 1.87, while on the right, the bars are
unlocked, with the Galactic bar at an OLR of 1.5 and the long bar with
an OLR of 1.87. Reducing the mass brings the simulations into much
better agreement with the observations, although the secondary feature
is at a too high $V$ velocity. This can be remedied by shifting to a
higher value of 1.90 for the OLR.

\section{Discussion \& conclusions}

We have performed simulations of the effects of bars in the inner Galaxy on the
distribution of stars in velocity space near the Sun. Our simulations are
similar to those of \cite{Dehnen2000} and \cite{Minchev07}, in which a large
library of orbits of stars passing through the Solar neighbourhood are
performed in a model of the Galactic potential including a bar. These earlier
studies showed that the Hercules stellar stream in the solar neighbourhood
could be the result of a resonance of stellar orbits with a fast moving bar
(i.e. with a co-rotation with the Galactic disc not much beyond the end of the
bar). Both \cite{Dehnen2000} and \cite{Minchev07} find that the Hercules stream
can be used to constrain the rotation rate and angle of the bar.

Our motivation to reexamine this is the recent discovery that the
Galaxy may contain a second bar, in addition to the traditional
Galactic bar, by \cite{Benjamin05} and \cite{Lopez07}. These authors
have found a `long bar' in the Galactic centre. It is quite different
to the Galactic bar, being highly flattened (semi-major axes of 3.9,
0.6 and 0.1 kpc as compared to 3.5, 1.4 and 1.0 kpc). Furthermore, it
is aligned at an angle of about 43 degrees to the line of sight to the
Galactic centre, as opposed to about 20 degrees for the Galactic bar.
For the Galactic bar we adopt a mass of $10^{10}$ M$_\odot$
\citep{Zhao96,Weiner99} and for the long bar 6 $\times$ 10$^9$
M$_\odot$ \citep{Lopez07}. The long bar has a mass quite comparable to
the traditional bar.

We have simulated the effects of the long bar, as well as the Galactic
bar, and both bars, on the velocities of stars in the Solar
neighbourhood. Rather than simulating bars as a quadrupole
perturbation in the local Galactic potential, we simulate the bars
with a Ferrers potential (which generates a triaxial ellipsoidal
density distribution). This is computationally more expensive, but
allows for more accurate modelling of a bar as a whole, especially for
stars passing through, or trapped in, a bar. The simulations are
performed in 2-D, although the model itself allows 3-D simulations.

We confirm the basic picture that the Hercules stream can be generated by a
resonance between local stars and the bar. Both the long bar and the Galactic
bar produce Hercules-like features in the Solar neighbourhood. The position of
the Hercules stream in the simulations is found to depend on the mass of the
bar, the rotation rate of the bar, and rather weakly, on the angle of the
bar. The geometry of the bars is not very important either, since both bars
generate Hercules-like streams.

We measure the rotation rate of the bars from the position of the Hercules
stream, finding that both bars can produce acceptable fits to of the $(U,V)$
velocities of nearby stars. We measure the rotation rate of both bars via the
Outer Lindblad Resonance value they produce (i.e. by the ratio of the period of
a star with velocity of the Local Standard of Rest in a given
Galactic model to the period of rotation of the bar). We measure
values of $1.87 \pm
0.02$ for the OLR of both the Galactic bar and the long bar. This
value gives a pattern speed of 55.9 and 54.9 km s$^{-1}$ kpc$^{-1}$,
respectively, for the Galactic bar and long bar. This puts the
corresponding co-rotation, for the long bar, as defined by its
potential model, at the approximate tip of the bar, 3.9 kpc. For the
Galactic bar, our model produces no co-rotation radius.
It should be noted
however that this value is sensitive to the assumed Solar motion. We assume the
Solar motion recently derived by \cite{Schonrich10}. Had we adopted the Solar
motion used by \cite{Dehnen2000}, we would have derived a value of $1.90 \pm
0.03$. Both of these derived values are in reasonable agreement with
\cite{Dehnen2000}'s determination of $1.85 \pm
0.15$ and \cite{Minchev07}'s determination of $1.87 \pm 0.04$.

After looking at the effects of each bar alone, we tried putting both bars into
the modelling. The effects on local velocity space are quite dramatic, as we
have now added quite a lot of mass in an non-axisymmetric component to the
Galactic centre. In particular, these simulations produce a striking tail of
stars in the Solar neighbourhood at high $U$ velocities, inconsistent with
observations. The simple expedient of halving the mass of both bars produces
much better fits to the data. On the basis of these experiments then, if there
are two bars in the Galactic centre, we wonder whether the mass estimates of
each bar include material from the other bar, and are perhaps overestimates.
This offers a challenge to observational studies of the bar masses.

Finally, we have found that an additional feature in local velocity
space can arise under some conditions, especially for the long bar
case.  A weak stream of stars can form at about $V\sim-100$ and $U<0$
km s$^{-1}$.
We believe this is due to a direct interaction of such stars with the
bar, rather than a resonance with the bar. It is tempting to associate
this feature with the known Arcturus stream at V$\sim-$100 km s$^{-1}$
\citep{Williams09} in the Solar neighbourhood. Thus, a dynamical
origin for the stream is possible, whereas people have mainly
discussed accretion origins (of a satellite galaxy) to date.

We point out that our method has shortcomings. Firstly, the
simulations are in 2-D, and while this is almost certainly acceptable
for typical disc stars with near circular orbits, full 3-D simulations
may be needed for features like that which we associate with the
Arcturus stream. In addition, there are no spiral arms or other
non-axisymmetric components, besides the bars, in the modelling, and
these may be responsible for additional complexity in local velocity
space which we cannot model. We are not affected by the problems of
long-term backwards integration as mentioned in \cite{Fux01}. Compared
to earlier studies, we try to more accurately represent the Galaxy's
observed properties, such as local densities and bar geometry. The
model also lets us look at direct interaction, as well as resonances,
this is especially important for high velocity stars (e.g. those in
the Arcturus stream), where they have high enough eccentricities to
enter the inner parts of the Galaxy. The model also makes it possible,
in the future, to fully study the 3D impact of the bar(s).

\section*{Acknowledgements}

The authors wish to thank Professor Kimmo Innanen for starting out this
project, as well as Professor Daniel Pfenniger for providing his code for
the bar potential. We also wish to thank Burkhard Fuchs and the referee for
insightful comments. EG acknowledges the financial
support of the Finnish Cultural Fund and the Finnish Graduate School in
Astronomy and Space Physics. CF acknowledges financial support by the
Academy of Finland.

\nocite{Galdyn,EFE,Carlson94}

\bibliographystyle{mn2e}
\bibliography{Gardner-Flynn}
\appendix
\label{lastpage}
\end{document}